\documentstyle[11pt]{article}
\begin{document}
\title{
\begin{flushright}
{\normalsize
IPS Report 95-13\\
hep-lat/9505021\\
}
\end{flushright}
{\bf Systematic errors of L\"uscher's fermion method\\ and its extensions}}

\author{Artan Bori\c{c}i and Philippe de Forcrand\\
        {\normalsize\it Interdisciplinary Project Center
                        for Supercomputing (IPS)}\\
        {\normalsize\it Swiss Federal Institute
                        of Technology Z\"urich (ETH)}\\
        {\normalsize\it IPS, ETH-Zentrum}\\
        {\normalsize\it CH-8092 Z\"urich}\\
        {\normalsize\it borici@ips.id.ethz.ch},
        {\normalsize\it forcrand@ips.id.ethz.ch}\\
}
\maketitle

\begin{abstract}
	We study the systematic errors of L\"uscher's formulation of
dynamical Wilson quarks and some of its variants, in the weak and strong
coupling limits, and on a sample of small configurations at finite $\beta$.
We confirm the existence of an optimal window in the cutoff parameter
$\varepsilon$, and the exponential decrease of the error with the number of
boson families. A non-hermitian variant improves the approximation further
and allows for an odd number of flavors. A simple and economical Metropolis
test is proposed, which makes the algorithm exact.
\end{abstract}

\section{Introduction}
	Hybrid Monte Carlo (HMC) is now the standard method to simulate
fermionic field theories \cite{HMC}.  It is exact, relatively simple
to implement, and its cost grows just a little faster than the volume $V$
of the system \cite{volume54}. This should not stop us from searching
for better algorithms.

	Fermionic interactions give rise, after integration of the fermionic
fields, to a non-local determinant. HMC addresses the problem of the
non-locality of this determinant by linearizing the action in a succession
of molecular dynamics
steps. The calculation of the force at each step is obtained via the iterative
solution of a sparse linear system. But this linearization implies
infinitesimal
steps, which in turn make narrow, very high energy barriers nearly impassable.
Thus very long autocorrelation times have been observed, eg. for the
topological charge \cite{KogutTeper}, and ergodicity may be questionable
when the vanishing of the fermion determinant divides phase space into
disconnected regions \cite{WhiteWilkins}. A finite step-size algorithm
is highly desirable.

	All such algorithms have been abandoned, because of their high
cost proportional to $V^2$ \cite{PhIOS}. Recently however, L\"uscher
proposed a finite-step bosonic formulation where one can control, through
the number of auxiliary bosonic fields, the trade-off between accuracy and
computer cost \cite{lusher1}. Surprisingly, no systematic investigation
of this trade-off has appeared in the literature, except in full-blown studies
of the complete QCD theory \cite{lusher2,t3d}. It is our purpose
to fill this gap here, by studying the weak and strong coupling limits,
and the interacting theory on a sample of HMC configurations.  In Section 2,
we recall L\"uscher's formulation of the QCD determinant, and explain our
methodology.  In Section 3, we present our results for the systematic error
of the method.  In Section 4, we investigate the effect of the usual
`even-odd splitting' of the Dirac operator. In Section 5, we propose a
non-hermitian variant, with improved convergence, suited also for an odd
number of quark flavors. Section 6 describes a simple Metropolis test which
makes the algorithm exact. Conclusions follow.

\section{L\"uscher's method}

	In cases where the fermionic determinant can be written
as a square $det Q^2$, L\"uscher has proposed a general method to approximate
it by a local bosonic action.
The essential steps are:

$\bullet$ find a real polynomial $P(x)$ of even degree $n$,
approaching $1/x$ as $n \rightarrow \infty$
over the spectrum of $Q^2$, which must be real.

$\bullet$ decompose $P(x)$ into a product of monomials
\begin{equation}
P(x) \equiv c_n x^n + . . . + c_1 x + c_0
     = c_{n} \; \prod_{k=1}^{n} (x - z_{k})
\end{equation}

\noindent
such that
\begin{equation}
det P(Q^2) = c_n^N \; \prod_{k=1}^{n}
             det (Q - \sqrt{\bar{z_{k}}}) det (Q - \sqrt{z_{k}})
\end{equation}

where $N$ is the rank of $Q^2$.

$\bullet$ express each factor above as a Gaussian integral over bosonic
fields, so that finally
\begin{equation}\label{gaussian}
det Q^2 \approx
\frac{1}{det P(Q^{2})} = \frac{1}{c_n^N (2{\pi}i)^n}
     \int \prod_{k=1}^{n} [d{\phi}_k^{\dag}] [d{\phi}_k]
     e^{-{\phi}_k^{\dag} (Q - \sqrt{z_{k}})^{\dag}
                         (Q - \sqrt{z_{k}}) {\phi}_k }
\end{equation}

Thus one trades the initial non-local determinant for a bosonic action
which is a sum of $n$ local terms. This bosonic action can be viewed as
a discrete path integral; it converges to the exact determinant
as the discretization step $\sim 1/n$ tends to zero.
Note that the normalization factor $c_n$ of the polynomial
drops out of any expectation value.

\bigskip
Here we study this approach when applied to dynamical Wilson fermions.
Let us denote the Dirac matrix by
\begin{equation}\label{Ddef}
D = ({\bf 1} - \kappa M)
\end{equation}

\noindent
where $M$ is the lattice Wilson hopping operator and
$\kappa$ the hopping parameter.
L\"uscher chooses for $Q$ the hermitian operator
\begin{equation}\label{Q_mat}
Q = c_0~\tilde{Q} \; ; \; \tilde{Q} = \gamma_{5} D
\end{equation}
where $c_0 = (1 + 8 \kappa)^{-1}$ is introduced to guarantee that the
norm of $Q$ is bounded by one,
and his method to approximate $det Q^2$ applies to 2 degenerate quark flavors.

The particular polynomial adopted by L\"uscher is built from
Chebyshev polynomials $T_n(x)$, with the definitions

\begin{equation}
T_{0}(x) = 1
\label{t0}
\end{equation}
\begin{equation}
T_{1}(x) = x = cos \theta
\label{t1}
\end{equation}
\begin{equation}
T_{n}(x) = cos(n\theta)
\label{tn}
\end{equation}

\noindent
The approximation polynomial is then defined by

\begin{equation}
P(x) = \frac{1 - R(x)}{x} = c_{n} \; \prod_{k=1}^{n} (x - z_{k})
\label{pol}
\end{equation}

\noindent
where $R(x)$ is the scaled and translated Chebyshev
polynomial such that $R(0) = 1$,

\begin{equation}
R(x) = \frac{T_{n+1}(  \frac{2 x}{1 - \varepsilon}
                     - \frac{1 + \varepsilon}{1 - \varepsilon})}
            {T_{n+1}(- \frac{1 + \varepsilon}{1 - \varepsilon})}
\label{Rx_def}
\end{equation}

\noindent
with $\varepsilon$ an adjustable parameter $\in (0,1)$.
By straightforward algebra, one verifies
that the zeroes of $P(x)$, $z_{k}, k = 1,...,n$ are given by

\begin{equation}\label{zeroes}
z_{k} = \frac{1 + \varepsilon}{2} (1 - \cos\frac{2 \pi k}{n+1})
             - i \sqrt{\varepsilon} \sin\frac{2 \pi k}{n+1}
\end{equation}

A very important advantage of using Chebyshev polynomials
is that the error of the approximation can be bounded,
and the bound converges exponentially with $n$:

\begin{equation}\label{l_bound}
| R(x) | = | 1 - x P(x) | \leq 2
\left(\frac{1 - \sqrt{\varepsilon}}
           {1 + \sqrt{\varepsilon}}\right)^{n+1},
\forall x \in [\varepsilon,1]
\end{equation}

\noindent
The proof is given in section 5.

\bigskip
\bigskip
Our methodology to assess the error of this method and to compare
it with other variants, is the following:
we compute the quantity

\begin{equation}
y \equiv det ~~Q^2 P(Q^2)
\label{y_def}
\end{equation}

\noindent
by calculating the eigenvalues of $Q^2$, $\lambda_{i}, i = 1, ... N$, so that

\begin{equation}
y = \prod_{i=1}^{N} \lambda_{i} P(\lambda_{i})
\end{equation}

\noindent
To properly cover the full range of gauge couplings,
we study the variations of $y$ in 3 different cases.

\bigskip
\noindent
i) For free fermions the spectrum of $Q$ is known analytically;
we estimate the error by monitoring
$|y^{1 / N} - 1|$.

\bigskip
\noindent
ii) In the strong coupling limit we
generate a sample of quenched $4^4$ configurations, and
measure the fluctuation of $y$ by monitoring

\begin{equation}\label{delta}
\Delta \equiv \frac{1}{<y>} \sqrt{<y^2> - <y>^2}
\label{Delta}
\end{equation}

\noindent
The reason for the normalization factor $1/<y>$ is to remove the dependence
of the error on the arbitrary constant $c_n$ (eq.(\ref{pol})).

\bigskip
\noindent
iii) At intermediate coupling, we perform the same analysis for a set of
matrices $Q^2$ generated
by hybrid Monte Carlo at $\beta = 6$ and $\kappa = 0.14$ on a $4^4$
lattice.

\section{Systematic errors of L\"uscher's method}

In the case of free fermions, our definition $|y^{1 / N} - 1|$ of the error
depends on the normalization coefficient $c_n$. A simple way to compute
$c_n$ relies on the observation that $R(\frac{1+\varepsilon}{2}) = 0$, so that
\begin{equation}
c_n^{-1} = \frac{1+\varepsilon}{2} \prod_{i=1}^{n}
          (\frac{1+\varepsilon}{2} - z_k)
\end{equation}
We then show in Fig.1a the error as a function of the parameter $\varepsilon$,
for 20, 54, 90 and 148 boson fields
($\kappa = 0.11$, on an $8^4$ lattice).
As expected, there is an optimal value
for $\varepsilon$: if $\varepsilon$ is too large, the polynomial approximation
degrades over the lower part of the spectrum of $Q^2$; if $\varepsilon$ is
too small, the approximation becomes poor over the whole spectrum, because
$(1 - \sqrt{\varepsilon})/(1 + \sqrt{\varepsilon})$ approaches 1
(see eq.(\ref{l_bound})). Not surprisingly, the optimal
value of $\varepsilon$ is slightly larger than the smallest eigenvalue of
$Q^2$, and approaches it as the number of bosonic fields increases.

At finite $\beta$, fluctuations of $\lambda_{min}(Q^2)$ complicate the matter:
configurations with small $\lambda_{min}(Q^2)$ contribute larger errors,
so that $\varepsilon$ must be tuned slightly ${\em below}$
$<\lambda_{min}(Q^2)>$
in order to minimize the average error, when this error is small.
This behaviour is shown in the (quenched) strong coupling
in Fig.1b for $\kappa = 0.2$. It remains unchanged at intermediate couplings
$\beta=6$ in Fig.1c, where we also illustrate
a typical spectrum of the matrix $D$ in Fig.2a.

{}From these figures,
it appears that the optimal ratio $\varepsilon / \lambda_{min}(Q^2)$ is
relatively insensitive to $\beta$. This should simplify the task of tuning
$\varepsilon$.

\subsection*{Rescaling the matrix}

As mentioned in \cite{lusher2}, rescaling the matrix $Q$
(see eq.(\ref{Q_mat})) to $Q/c_M$ can be useful.
This is equivalent to rescaling
the roots of the polynomial $z_k \rightarrow c_M^2 \; z_k$.
In \cite{lusher2} one assumes $c_M \geq 1$ to guarantee
that the spectrum is bounded by one. Nonetheless,
one may let $c_M$  be less than one,
if one monitors the largest eigenvalue of $Q^2$.
Under rescaling, this largest eigenvalue approaches 1,
making use of the full interval $[\varepsilon,1]$ of validity of the
approximation (\ref{gaussian}). Simultaneously the smallest eigenvalue
of $Q^2$ increases, with corresponding gains in convergence. One
readily sees from equation (\ref{l_bound}) that the number $n$ of boson
fields can be multiplied by $\approx c_M$.

The benefits of tuning $c_M$ vanish in the weak coupling limit, since
the largest eigenvalue of $Q^2$ then tends to 1.
Even in the strong coupling, the advantage remains small, ${\cal O}(10-20)\%$.
But the normalization of $Q$ must be considered carefully when one uses
even-odd splitting of the lattice sites.

\section{Even-odd splitting}

L\"uscher's method is based on a polynomial
approximation of the inverse, just like iterative methods to find
the quark propagator, with the difference that the polynomial
is {\em predetermined} in L\"uscher's case, whereas it is built recursively
and {\em adaptively} by an iterative linear solver.
The similarity of the two approaches should encourage us to
try here what works well there.
In this section we test the simplest such idea, that of partitioning
the lattice sites into `even' and `odd', and of factorizing $det Q^2$
into equal, even and odd factors.

To justify such a factorization,
we simplify the derivation used in \cite{mart-ot} for staggered fermions.
Define a diagonal operator $\Sigma$ with entries $1$ on even sites
and $-1$ on odd sites.
This operator anticommutes with the
hopping matrix $M$, since $M$ connects even and odd sites.
Noting that ${\Sigma}^2 = {\bf 1}$, one gets

\begin{equation}\label{17}
det D = det ({\bf 1} - \kappa M) = det \Sigma ({\bf 1} + \kappa M) \Sigma
= det ({\bf 1} + \kappa M)
\end{equation}

\noindent
Now $D^\dagger = \gamma_5 D \gamma_5$, and $\gamma_5^2 = {\bf 1}$, so that

\begin{equation}
det \tilde{Q}^2 = (det D)^2 = det ({\bf 1} - {\kappa}^2 M^2)
\end{equation}

\noindent
Furthermore we observe that $M^2$ commutes with $\Sigma$; then if $\vec{x}$
is an eigenvector of $M^2$, so is $\Sigma \vec{x}$, or
$({\bf 1} \pm \Sigma) \vec{x}$, which is non-zero on even or odd sites
respectively. Therefore all eigenvalues are two-fold degenerate, and one has

\begin{equation}
  det ({\bf 1} - {\kappa}^2 M^2)_{even}
= det ({\bf 1} - {\kappa}^2 M^2)_{odd}
\end{equation}

\noindent
So we can finally write

\begin{equation}\label{20}
det \tilde{Q}^2 = det ({\bf 1} - {\kappa}^2 M^2)_{even}^2
\label{eo}
\end{equation}

\noindent
This way we can work with bosonic fields defined on
even sites only, saving a factor 2 in memory. The bosonic action
however is now less local, so that a bosonic update requires about as many
operations as before. Nonetheless, further gain may come through a reduction
in the number of bosonic families required for a given accuracy.

The reason for this potential gain becomes clear when one looks at the
spectrum of a typical configuration: the spectrum of Fig.2a shrinks to
that of Fig.2b after even-odd preconditioning. One can now readjust
$\varepsilon$ and vary the number $n$ of bosonic families,
with and without even-odd splitting. In both cases,
the error decreases exponentially with $n$, as predicted by eq.(\ref{l_bound});
but it does so much faster in the even-odd splitting formulation,
as illustrated in Figs. 3a and 3b for $\beta=0$ and $\beta=6$ respectively.
In both figures the gain is a factor 2 to 3 in $n$.

This large gain can be understood by considering the spectral
radius of $\kappa M$, say $\rho \equiv 1 - \gamma$: the smallest eigenvalue of
entering the determinant eq.(\ref{17}) is $\gamma \ll 1$, but that in
eq.(\ref{20})
is $\sim 2 \gamma$. Thus the optimal value of $\sqrt{\varepsilon}$ should
also increase by a factor 2, which according to eq.(\ref{l_bound}) allows
a reduction of $n$ by the same factor. In addition, one can achieve
important gains by tuning the rescaling constant $c_M$ to bring the
largest eigenvalue of $Q^2$ close to 1, as explained in the previous
section. The normalization constant which guarantees that the norm of
$\gamma_5 ({\bf 1} - {\kappa}^2 M^2)_{even}$ is bounded by one is now
$c_0 = (1 + 64 \kappa^2)^{-1}$. But with this normalization, the largest
eigenvalue of $Q^2$ becomes rather small away from $\beta = \infty$.
Thus in Figs.3a and 3b both, we tuned $c_M$ to 0.6 instead
of its default value of 1.

\section{Non-hermitian variant}

\subsection{Formulation}

Another lesson can be learned from iterative solvers \cite{us}:
the method of biconjugate gradients (BiCG) \cite{lanczos2} requires less
work than that of conjugate gradients (CG) \cite{hestens}.
BiCG applies Lanczos polynomials on the matrices $D$
and $D^{\dagger}$ to construct the solution,
whereas CG applies the Lanczos polynomial on the matrix $Q^{2}$.
This observation motivated us to look for an approximation to $1/D$ itself,
instead of $1/Q^2$. This should be possible as long as $det D$ is positive.
But a negative determinant can only be caused by negative {\em real}
eigenvalues; this situation can only occur for very small quark masses
as studied in the quenched case by \cite{Iwasaki},
beyond those reachable by present-day simulations of full QCD.
The benefits of the approach proposed below are two-fold: convergence
of the approximation is improved; and the simulation of an odd number
of quark flavors (or of any number of non-degenerate flavors) becomes
possible.

Let $P(z)$ be a polynomial of even degree $n$, with real coefficients,
defined in the complex plane (since the spectrum of $D$ is complex),
with complex conjugate roots $z_k, Im z_k \neq 0, k = 1,...,n$.
Assume that this polynomial satisfies $P(z) \rightarrow 1/z$ for
$n \rightarrow \infty, \forall z \in {\cal S}$, where ${\cal S}$ is
a domain in the complex plane containing the spectrum of $D$.
Then one can write

\begin{equation}
det P(D) = c_{n}^N \; \prod_{k=1}^{n/2}
           det (D - \bar{z_{k}}) det (D - z_{k})
\end{equation}

\noindent
Using the fact that $D = \gamma_5 D^{\dagger} \gamma_5$
and $\gamma_5^2 = {\bf 1}$ one gets

\begin{equation}
det (D - \bar{z_{k}}) = det (D^{\dagger} - \bar{z_{k}})
\end{equation}

\noindent
It follows that

\begin{equation}
det P(D) = c_{n} \; \prod_{k=1}^{n/2}
                det (D - z_{k})^{\dagger} det (D - z_{k})
\end{equation}

\noindent
and, in analogy with eq.(\ref{gaussian})
\begin{equation}
det D \approx
\frac{1}{det P(D)} = \frac{1}{c_{n}^N (2{\pi}i)^{n/2}}
     \int \prod_{k=1}^{n/2} [d{\phi}_k^{\dag}] [d{\phi}_k]
     e^{-{\phi}_k^{\dag} (D - z_{k})^{\dag}
                         (D - z_{k}) {\phi}_k }
\end{equation}

The approximation polynomial can be defined as before by

\begin{equation}
P(z) = \frac{1 - R(z)}{z} = c_n \; \prod_{k=1}^{n} (z - z_{k})
\end{equation}

\noindent
where $R(z)$ is a polynomial of degree $n+1$ such that $R(0) = 1$.
$|R(z)|$ should be small in the domain ${\cal S}$ of the approximation.
The spectrum of $D$ has a more or less elliptical shape, as can be seen
from Fig.2a. It is then natural to choose for the boundary
${\partial\cal S}$ an ellipse.
In that case (provided the origin is not inside the ellipse), it turns
out that the polynomial which minimizes $max_{\cal S} |R(z)|$ is again
a Chebyshev polynomial \cite{mant}.

The same definitions eqs.(\ref{t0}-\ref{tn}) remain valid for $T_n(w)$,
where now $w = x + iy$ and $\theta = \theta_{1} + i \theta_{2}$, that is

\begin{equation}\label{cos_expan}
w = cos\theta
  = cos\theta_{1}~cosh\theta_{2} - i~sin\theta_{1}~sinh\theta_{2}
\end{equation}

\noindent
where $\theta_1 \in [-\pi,\pi]$ and, for instance,
$\theta_2 \in [0,+\infty[$.  Note that, for $\theta_2$ fixed,
$w(\theta_1)$ describes an ellipse in the complex plane, with foci $\pm 1$
and semiaxes $cosh\theta_{2}$ and $sinh\theta_{2}$.

\noindent
{}From these definitions we can now specify the form of $R(z)$.
Choose for $\partial{\cal S}$ the ellipse centered at $(d,0)$, with large
semiaxis $a < d$, and focal distance $c \le a$.
Then the scaled and translated Chebyshev polynomial $R(z)$ is defined by

\begin{equation}
R(z) = \frac{T_{n+1}(\frac{z - d}{c})}{T_{n+1}(- \frac{d}{c})}
\end{equation}

\noindent
{}From this definition one can show, in analogy with the hermitian case,
that the roots of $P(z)$, $z_{k}, k = 1,...,n$ are

\begin{equation}
z_{k} = d (1 - \cos\frac{2 \pi k}{n+1}) - i \sqrt{d^2 - c^2}
               \sin\frac{2 \pi k}{n+1}
\end{equation}

\noindent
so that the $z_k$'s lie on the ellipse of same center and foci
as $\partial{\cal S}$, which goes through the origin.

We can now derive an error bound for the above polynomial
approximation in the complex plane. It states that

\begin{equation}\label{zbound}
| R(z) | = | 1 - z P(z) | \leq
2~\left( \frac{a + \sqrt{a^2 - c^2}}{d + \sqrt{d^2 - c^2}} \right)^{n+1},
{}~\forall z \in {\cal S}
\end{equation}

{\em Proof.} By definition

\begin{equation}
R(z) = \frac{\cos(n+1)\theta}{\cos(n+1)\theta_{0}}
\end{equation}
where
\begin{equation}
\cos\theta = \frac{z - d}{c}, \; \cos\theta_{0} = - \frac{d}{c}
\label{def}
\end{equation}

\noindent
{}From eq.(\ref{cos_expan}), one sees that $\theta_0 = \pi + i \alpha$,
where $\alpha \equiv cosh^{-1} d/c$.
Writing $\theta = \theta_1 + i \theta_2$, one gets

\begin{equation}
R(z) = \frac{-1}{cosh (n+1) \alpha} ( cos (n+1)\theta_1~cosh (n+1)\theta_2
     - i~sin(n+1)\theta_1~sinh (n+1)\theta_2 )
\end{equation}

\noindent
so that

\begin{equation}
|R(z)| \leq~\frac{cosh (n+1)\theta_2}{cosh (n+1)\alpha}
\end{equation}

\noindent
The coordinates of any point $z$ in ${\cal S}$ can be expressed as
$(\theta_1,\theta_2)$. Successive $\theta_2$'s define nested ellipses,
all of center $d$ and focal distance $c$: the innermost one, for
$\theta_2 = 0$, is the real segment $[d - c, d + c]$; the outermost one,
for $\theta_2 = \theta_{max} \equiv cosh^{-1} a/c$, is ${\partial\cal S}$.
Along each such ellipse $|R(z)|$ is bounded by
$\frac{cosh (n+1)\theta_2}{cosh (n+1)\alpha}$,
and reaches this bound when $z$ is real.
The bound increases with $\theta_2$, so that over ${\cal S}$,
$|R(z)|$ is bounded by

\begin{equation}
\frac{cosh (n+1)\theta_{max}}{cosh (n+1)\alpha} =
e^{(n+1) (\theta_{max} - \alpha)}
             ~\frac{1 + e^{- 2 (n+1) \theta_{max}}}
                   {1 + e^{- 2 (n+1) \alpha}}
\end{equation}

The second factor is bounded by 2. Substituting $\theta_{max}$ and $\alpha$
by their definitions in terms of $a, c, d$, one recovers eq.(\ref{zbound}).

This derivation makes it clear that the error $|R(z)|$ is maximum on the
real points of $\partial{\cal S}$, but decreases exponentially inside
${\cal S}$. This is in sharp contrast with the hermitian approximation
of L\"uscher.

\subsection{Optimization and comparison with the hermitian case}

Two limiting cases for $\partial{\cal S}$ are of special interest:
the circle ($c=0$), which corresponds to the spectrum of $D$ in the
(quenched) strong coupling; and the real line segment $[d - a, d + a]$ ($c=a$),
which corresponds to the spectrum of $Q^2$ originally considered by L\"uscher.
For these 2 cases the error bound eq.(\ref{zbound}) becomes
$2 (\frac{a}{d})^{n+1}$ and $2 (\frac{a}{d + \sqrt{d^2 - a^2}})^{n+1}$
respectively. To make contact with the hermitian error bound
eq.(\ref{l_bound}), we just express
$a = \frac{1 - \varepsilon}{2}$ and $d = \frac{1 + \varepsilon}{2}$.
We recover then eq.(\ref{l_bound}) over the real segment $[\varepsilon,1]$;
over the circle we get:

\begin{equation}\label{u_bound}
|R(z)| \leq 2~\left( \frac{1 - \varepsilon}{1 + \varepsilon} \right)^{n+1}
\end{equation}

\noindent
We are now in a position to compare the hermitian and non-hermitian
approximations, in the strong coupling limit.
Call $\eta \sim m_{quark}$ the smallest eigenvalue of $D$.
Then the smallest eigenvalue of $Q^2$
will be ${\eta}^2$, and the hermitian and non-hermitian bounds
eqs.(\ref{l_bound}) and (\ref{u_bound}) are identical\footnote{
We greatly simplify here a subtle issue: if the smallest {\em singular}
value of $D$ is $\eta$, then the smallest eigenvalue of $Q^2$ is $\eta^2$;
but the relationship between {\em eigen}values of $D$ and of $Q^2$ could be
quite different, especially in the chirally broken phase. In particular
one easily proves that $\sigma_{min}(D) \le |\lambda|_{min}(D)$,
so the comparison we make here is a worst case scenario: the non-hermitian
variant will always be at least as accurate as the hermitian one.}.
It is true that only $n/2$ bosonic fields are necessary to generate
$det(D)$ in the non-hermitian case; but for 2 degenerate quark flavors,
one needs another $n/2$ fields to finally approximate $det(D)^2$,
for the same total of $n$ fields as in the hermitian case.

When $\beta$ increases, the error bound for the non-hermitian case
improves. The spectrum of $D$ can then be contained in a more elongated ellipse
whose aspect ratio tends to 2 in the free field limit.
It is straightforward to verify from eq.(\ref{zbound}) that the
convergence rate is multiplied by 2, allowing a reduction of $n$
by the same factor.

As importantly, the error in the non-hermitian approximation decreases
exponentially inside ${\cal S}$. To exhibit this difference with the
hermitian approximation, we show in Fig.4 the magnitude of the error
along the real axis,
for $\eta = 0.1$ and $n = 20$. That is, the range of the approximation
in the hermitian case is $[0.01,1]$, whereas in the non-hermitian
case it is $[0.1,1]$.
In the hermitian case, the error oscillates with constant amplitude
${\cal O}(10^{-2})$ over the interval $[\eta^2,1]$.
In the non-hermitian case, when $\partial{\cal S}$ is a circle
the error at $x = \eta$ and $x = 1$ is also ${\cal O}(10^{-2})$;
when $\partial{\cal S}$
is an ellipse of aspect ratio 2, the error  at $x = \eta$ and $x = 1$
is about squared. Either way, the error falls off exponentially inside
the approximation range, until it becomes uniformly oscillating in the
segment $[d-c,d+c]$.
So the accuracy on interior eigenvalues of $D$ is much improved.

We make numerical tests of this non-hermitian approximation
for HMC and strong coupling configurations.
We apply the same methodology as before,
but in the non-hermitian case we measure the fluctuations of
the quantity

\begin{equation}
y = \prod_{i=1}^{N} (\lambda_{i} P(\lambda_{i}))^2
\end{equation}

\noindent
where now $\lambda_{i}, i=1,...,N$ are eigenvalues of the quark
matrix $D$. With this definition of $y$ we can compare the
results directly to the hermitian case. In Figs.5
we show that the errors of the non-hermitian approximation
are reduced faster than in the hermitian approximation.
{}From Fig.5a we can see that in
the quenched strong coupling, the gain is a factor 4 to 5,
increasing as the quark mass is reduced. This large gain is
caused by the high density of interior eigenvalues.
In Fig.5b at $\beta = 6$, the gain is a factor $\sim 1.5$,
coming mostly from the elliptical shape of ${\partial\cal S}$.
Note that simulating one flavor to the same accuracy would require
half as many fields.

\section{Metropolis test}

	L\"uscher's original proposal includes the monitoring of the
error, and the possibility of obtaining exact results by re-weighting
the Monte Carlo measurements of each observable. The natural way to
calculate the error eq.(\ref{y_def}) is to express it in an eigenbasis of
$Q^2$,
obtained by the Lanczos algorithm. However two obstacles appear:
the Lanczos algorithm is affected by roundoff errors, which can be
controlled only if eigenvalue multiplicities are known \cite{Cullum},
and its cost grows like the square of the volume $V$ of the lattice,
so that it becomes overwhelmingly expensive on large lattices.
Nevertheless attempts at using the Lanczos method in a Metropolis test
show that most if not all of the error can be removed \cite{peardon,t3d}.

	In fact, it is sufficient to construct an unbiased estimator
of the ratio of errors between the new and the old configurations.
One can then use the noisy Monte Carlo method of \cite{KutiKennedy}, which
was successfully applied to fermionic simulations before the advent of
Hybrid Monte Carlo \cite{PhIOS,Pendleton}.

	The ratio to estimate is $det( D' P(D') )^2 / det( D P(D) )^2$,
calling $D'$ and $D$ the Dirac operators for the new and old configurations
respectively.
Taking the denominator as a partition function, this ratio can be rewritten
\begin{equation}
	< e^{- \eta^{\dagger} ( W^{\dagger} W - 1 ) \eta} >
\end{equation}
where the average $< >$ is taken over all Gaussian vectors $\eta$, and
$W = [D' P(D')]^{-1} D P(D)$. It is sufficient to estimate this ratio by
taking one Gaussian $\eta$ only. This requires the solution of a linear
system, which will take just a few iterations since the matrix $D' P(D')$
is almost the identity. The cost of this additional step is therefore
similar to that of an update of the $\phi_k$'s. As the volume $V$ of the
lattice grows, one should keep constant the acceptance of this Metropolis
test. This is achieved if the error per eigenvalue scales like $V^{-1}$.
Because the approximation is exponential in $n$, this can be accomplished by
a modest increase $\propto Log V$ in $n$,
and a CPU cost scaling like $V (Log V)^2$. Under this rescaling of $V$,
the number
of iterations necessary to solve the linear system above remains constant,
so that the overhead of the Metropolis test, measured in update sweeps,
does not change. Similarly, when the quark mass is decreased but the
acceptance is kept constant, the overhead of the Metropolis step, measured
in update sweeps, will not change.

\section{Conclusion}

	We have studied the systematic errors of L\"uscher's method to
simulate dynamical quarks, in the full range of strong to weak coupling.
Two parameters define the approximation, $\varepsilon$ and $n$:
$\varepsilon$ is the lower cutoff of the polynomial approximation,
$n$ is the number of auxiliary bosonic families.
We confirm the importance of tuning $\varepsilon$ near the
minimum eigenvalue of $Q^2$; the optimal ratio
$\varepsilon / <\lambda_{min}(Q^2)>$ approaches 1 from above as $n$ increases,
in a manner almost independent of $\beta$; it actually becomes slightly less
than 1 for large $n$, because configurations with small $\lambda_{min}(Q^2)$
give larger errors. For $\varepsilon$ fixed, we
confirm the exponential convergence of the approximation with $n$.

	In addition, we have studied the improvement of even-odd
preconditioning: it allows a reduction of $n$ by a factor 2 to 3.
A full reduction however can only be achieved by tuning to less than 1
the normalization parameter $c_M$ of the matrix $Q$.

	We have introduced a non-hermitian variant, which allows
the simulation of an odd number of quark flavors. The number of bosonic
families is proportional to that of flavors. For 2 flavors, the
non-hermitian formulation allows a reduction of $n$ by a factor
1.5 to 5. Full Monte Carlo
tests of this variant are in progress. They will shed some light on
the critical dynamics of the bosonic fields $\phi_k$, which depend on the
singular values of $(D - z_k)$, and might be different from the original
hermitian version.

	Even-odd preconditioning and non-hermitian formulation are most
advantageous for large quark masses and small $\beta$ respectively.
These two improvements can be combined when appropriate.

	Finally we have proposed a Metropolis test which removes any
approximation.  The overhead of this Metropolis test is modest and
independent of the lattice volume and the quark mass.

	The error we have considered is that on the fermionic determinant
itself. The error measured on a given observable during a Monte Carlo
simulation of the L\"uscher action will depend on the overlap of that
observable with the various eigenmodes of the determinant, and may be
smaller than we measured here.

\section*{Acknowledgements}
A.B. and Ph. de F. would like to thank Klaus Gaertner and Andrea Galli
respectively for discussions.

\newpage
{\Large \bf Figures:}

\begin{itemize}

\item
Figure 1: the magnitude of the error on the determinant as a function of
$\varepsilon$, for $n = 20, 54, 90, 148$, in the case of free field (Fig.1a),
(quenched) strong coupling (Fig.1b), and at $\beta = 6$ (Fig.1c).
$\kappa = 0.11,~0.2,~0.14$ respectively.
Arrows mark the minimum and the average of $\lambda_{min}(Q^2)$ over the
ensemble of configurations.

\item
Figure 2: a typical spectrum of the Dirac matrix $D$, eq.(\ref{Ddef}), in the
complex plane ($4^4$ lattice, $\beta = 6$, $\kappa = 0.14$). In Fig.2b
the spectrum is shown after even-odd preconditioning.

\item
Figure 3: the magnitude of the error on the determinant as a function of
$n$, in the original formulation (dotted line) and after even-odd
preconditioning (solid line). Fig.3a corresponds to $\beta = 0, \kappa = 0.2$,
Fig.3b to $\beta = 6, \kappa = 0.14$.

\item
Figure 4: the magnitude of the error of the polynomial approximation $P(x)$
as $x$ varies from 0 to 1. The three approximations shown are the original
hermitian approximation of L\"uscher, the non-hermitian approximation
inside a circle as appropriate for $\beta = 0$, and the non-hermitian
approximation inside an ellipse of aspect ratio 2, as appropriate for
$\beta = \infty$. All cases correspond to the same quark mass.

\item
Figure 5: the magnitude of the error on the determinant as a function of
$n$, in the original formulation (dotted line) and in the non-hermitian
variant (solid line). Fig.5a corresponds to $\beta = 0$, Fig.5b to
$\beta = 6$.

\end{itemize}

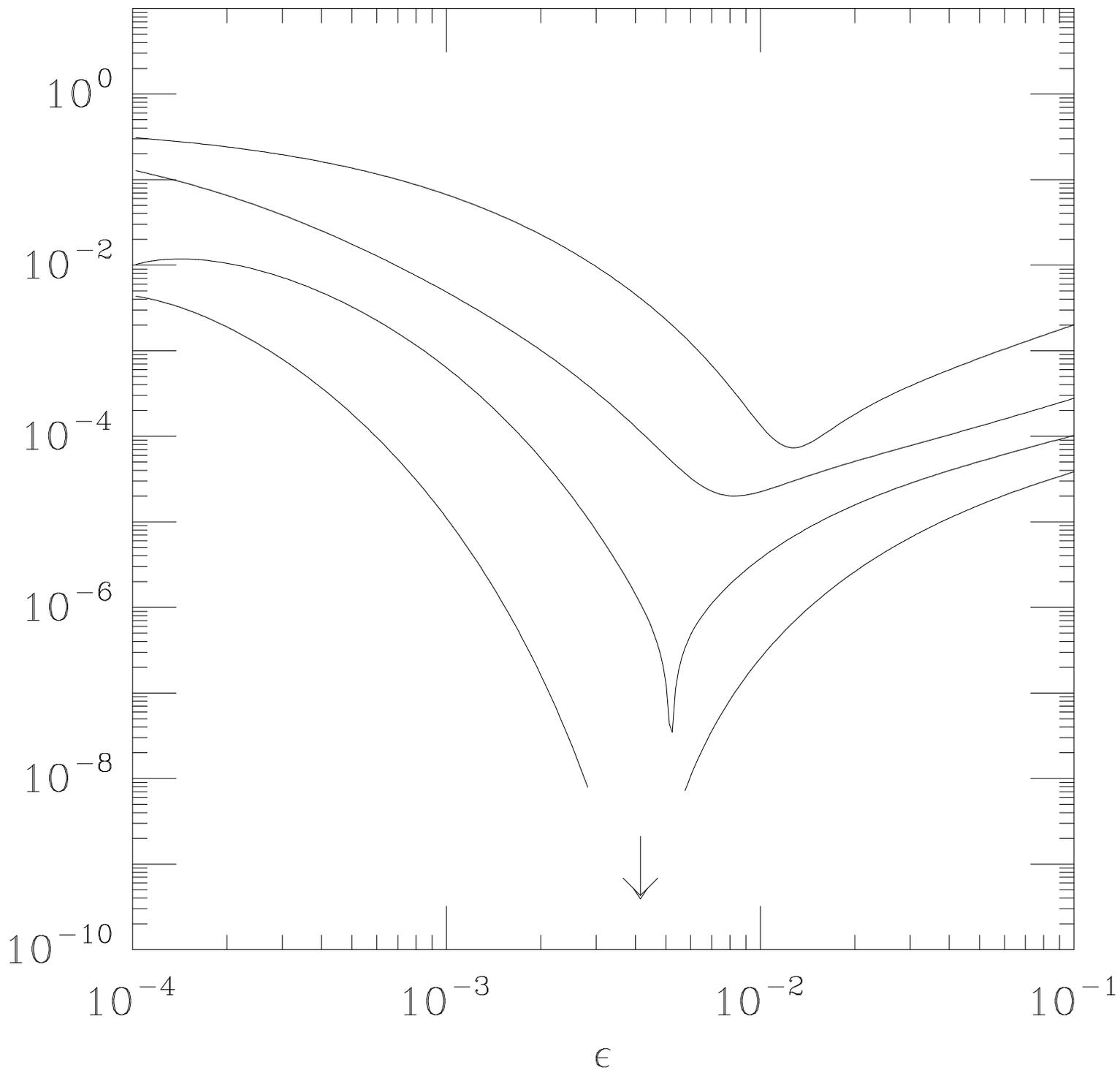
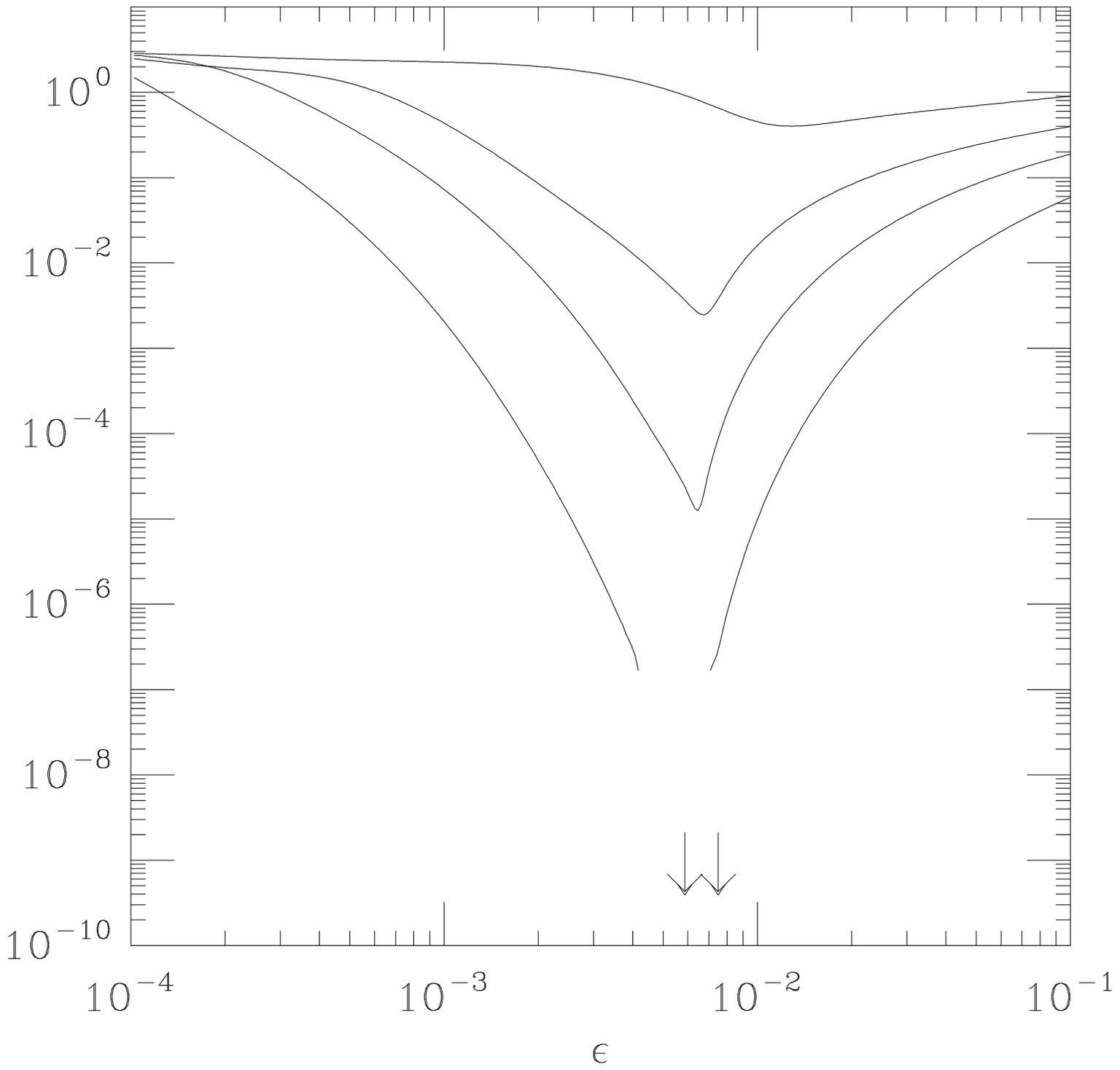
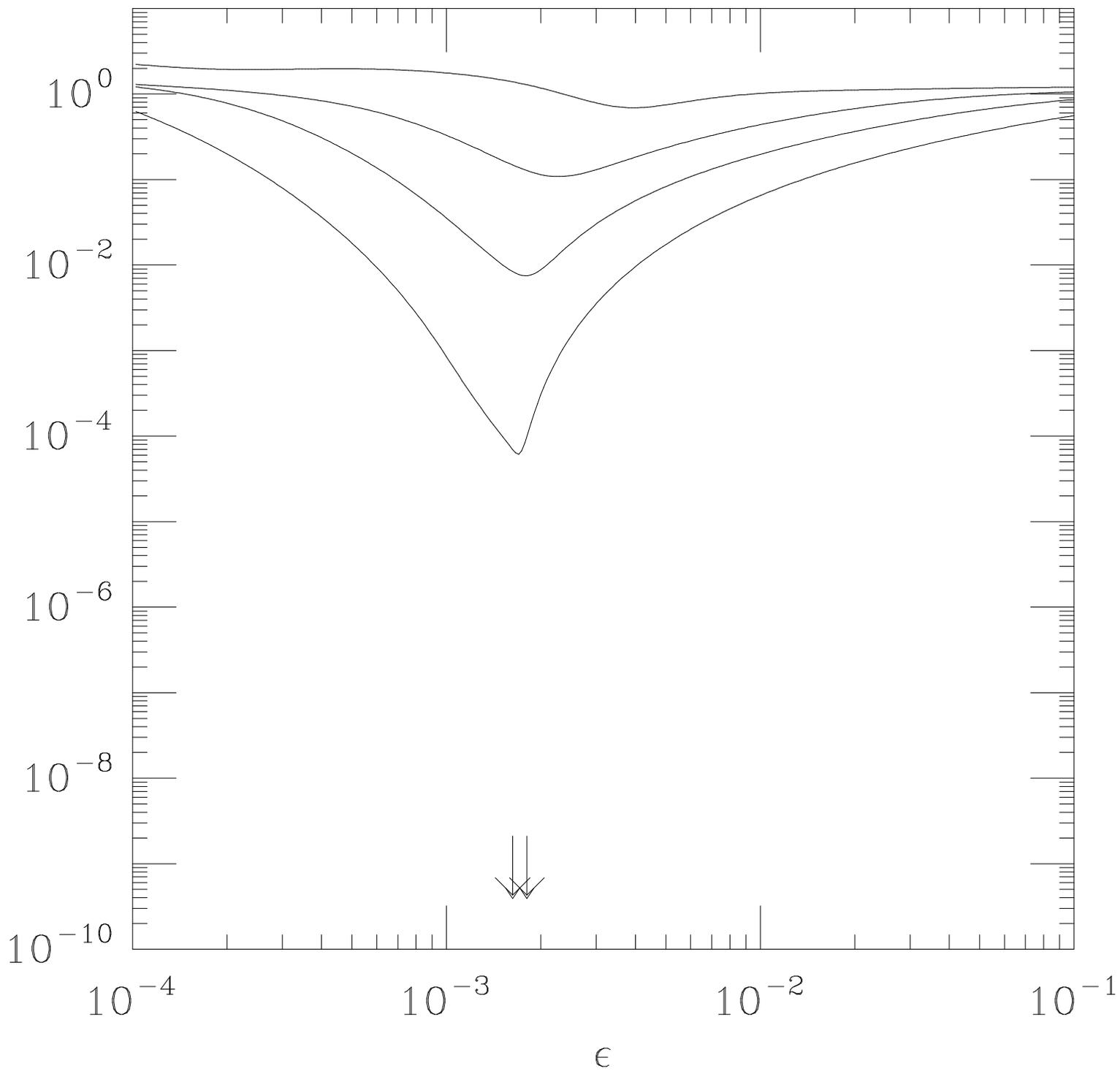
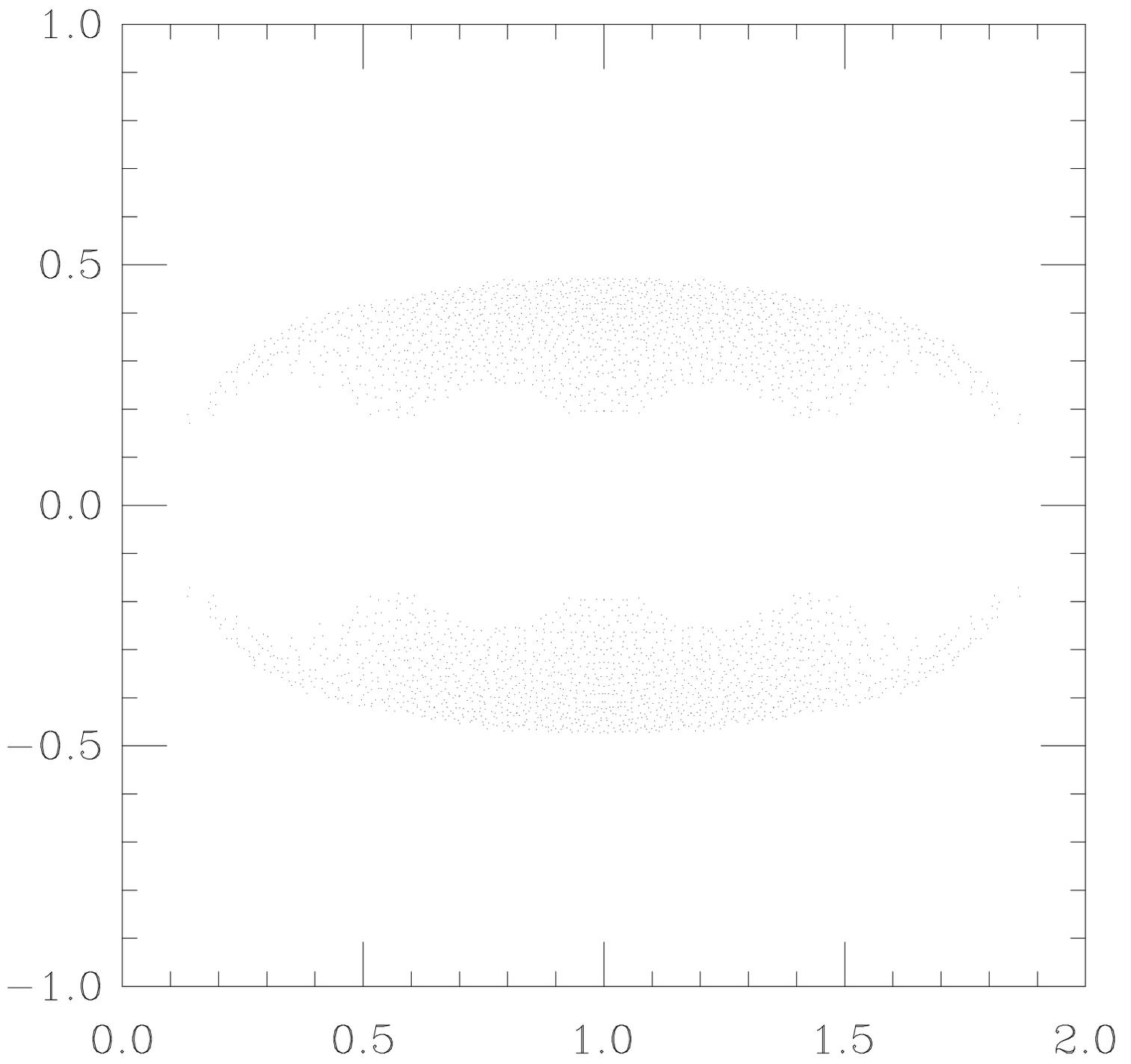
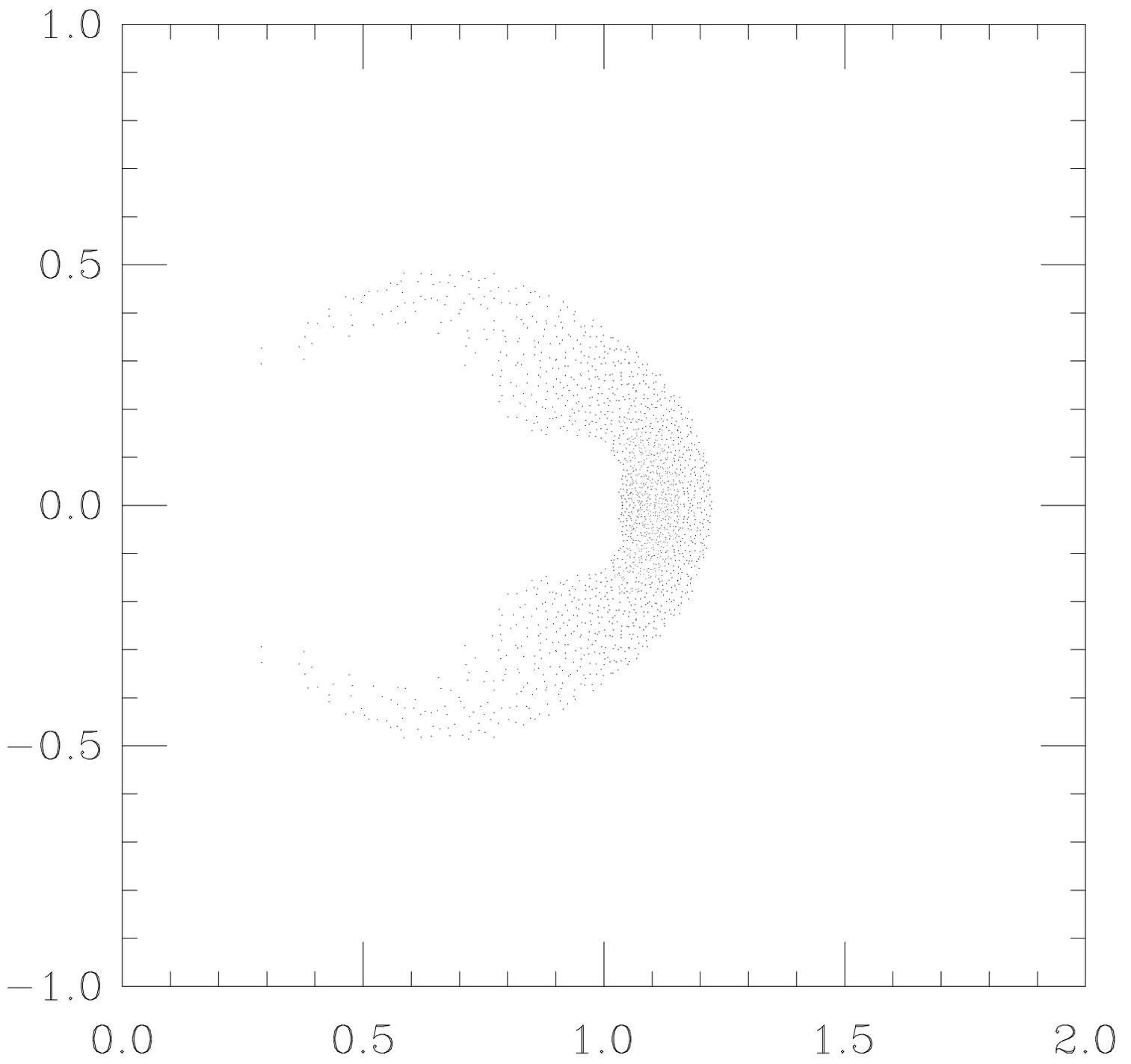
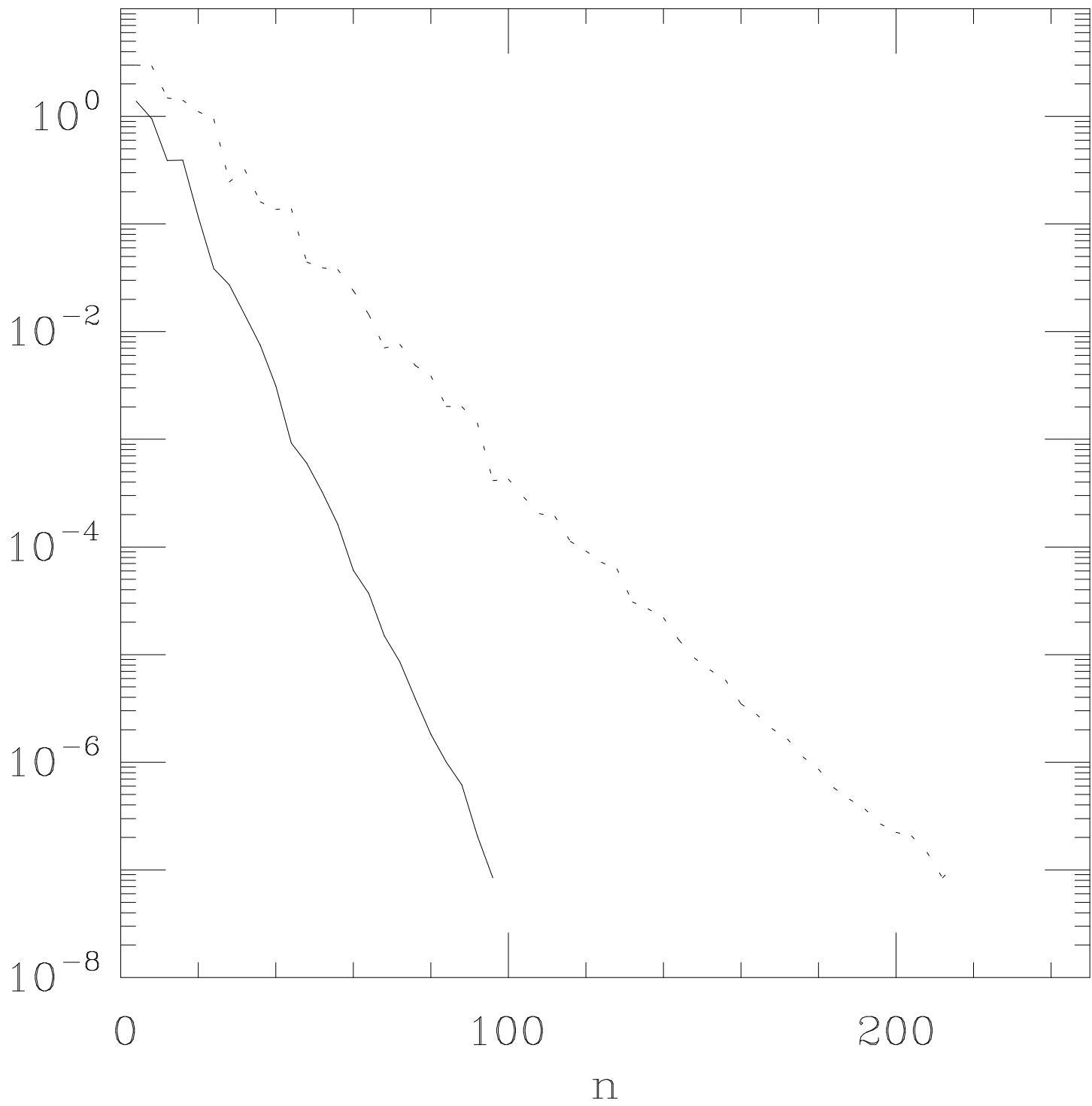
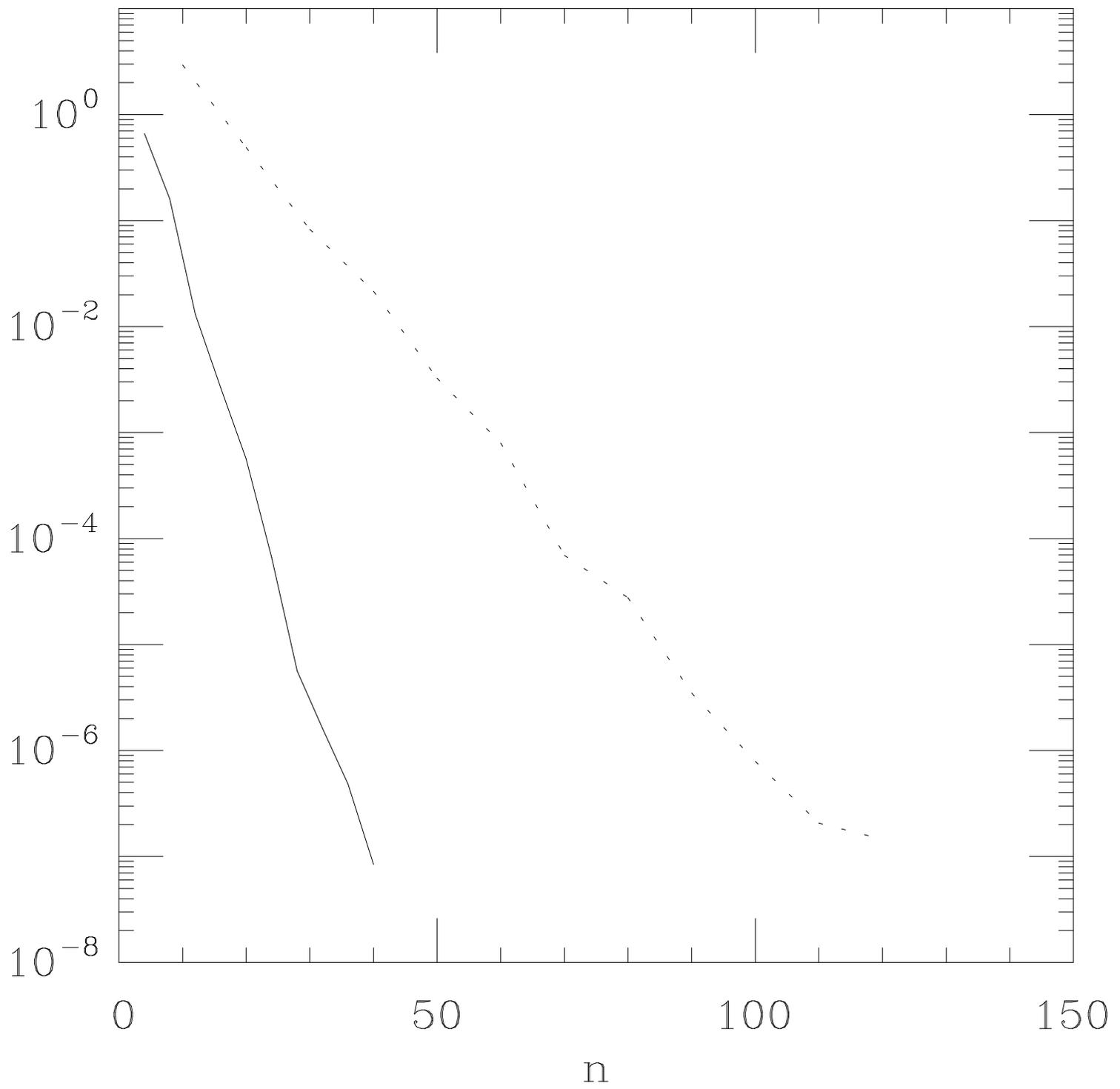
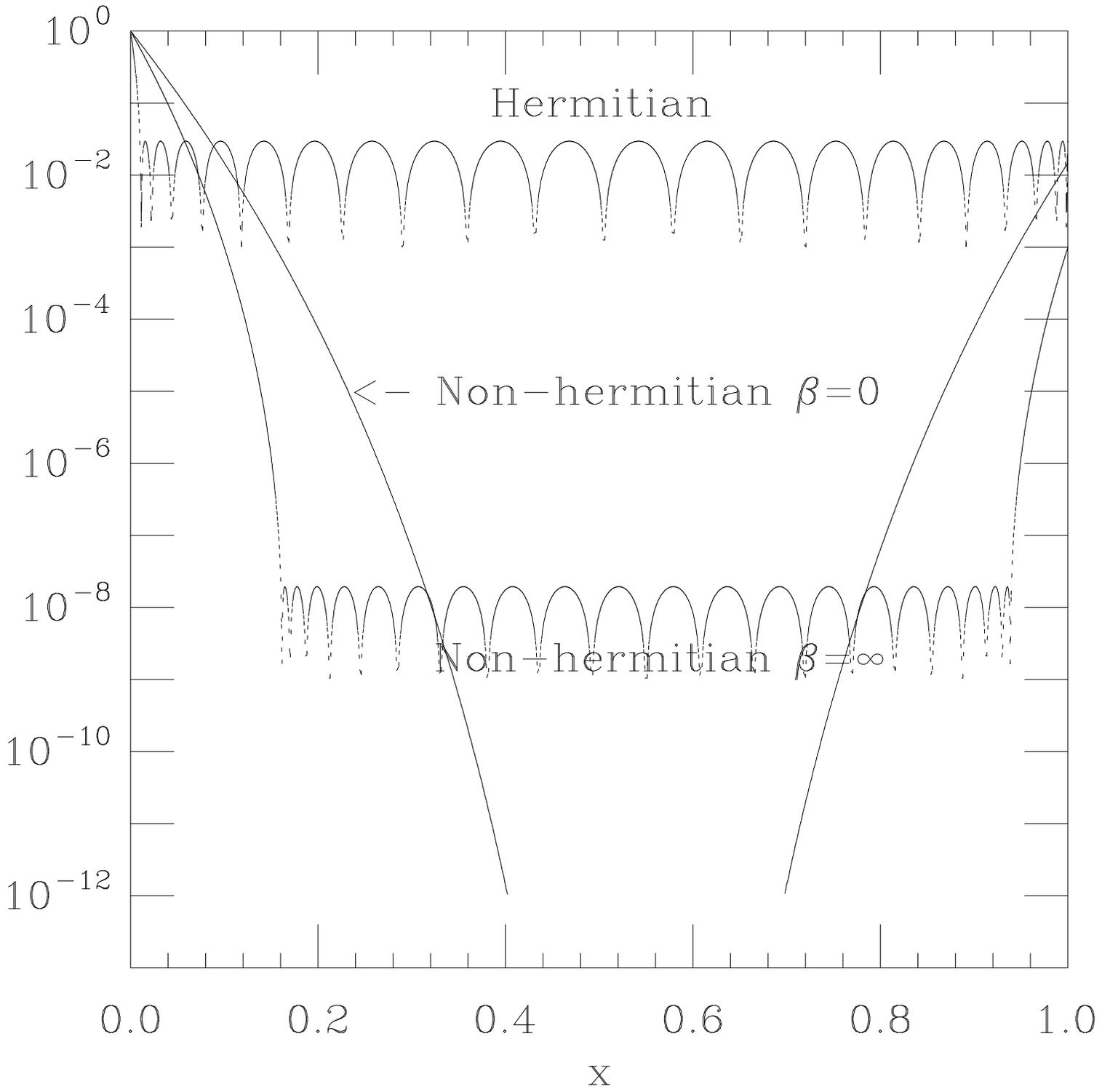
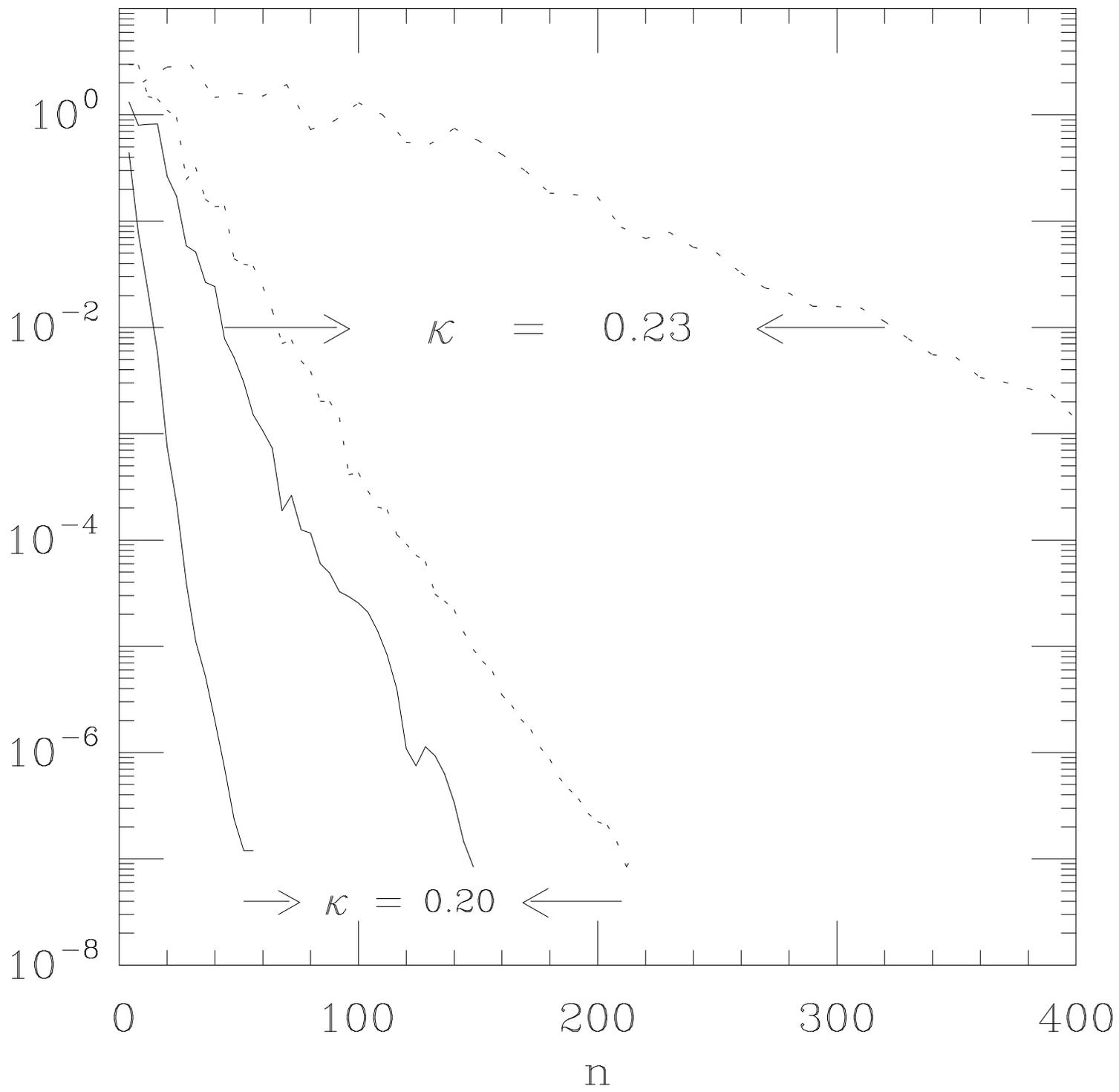
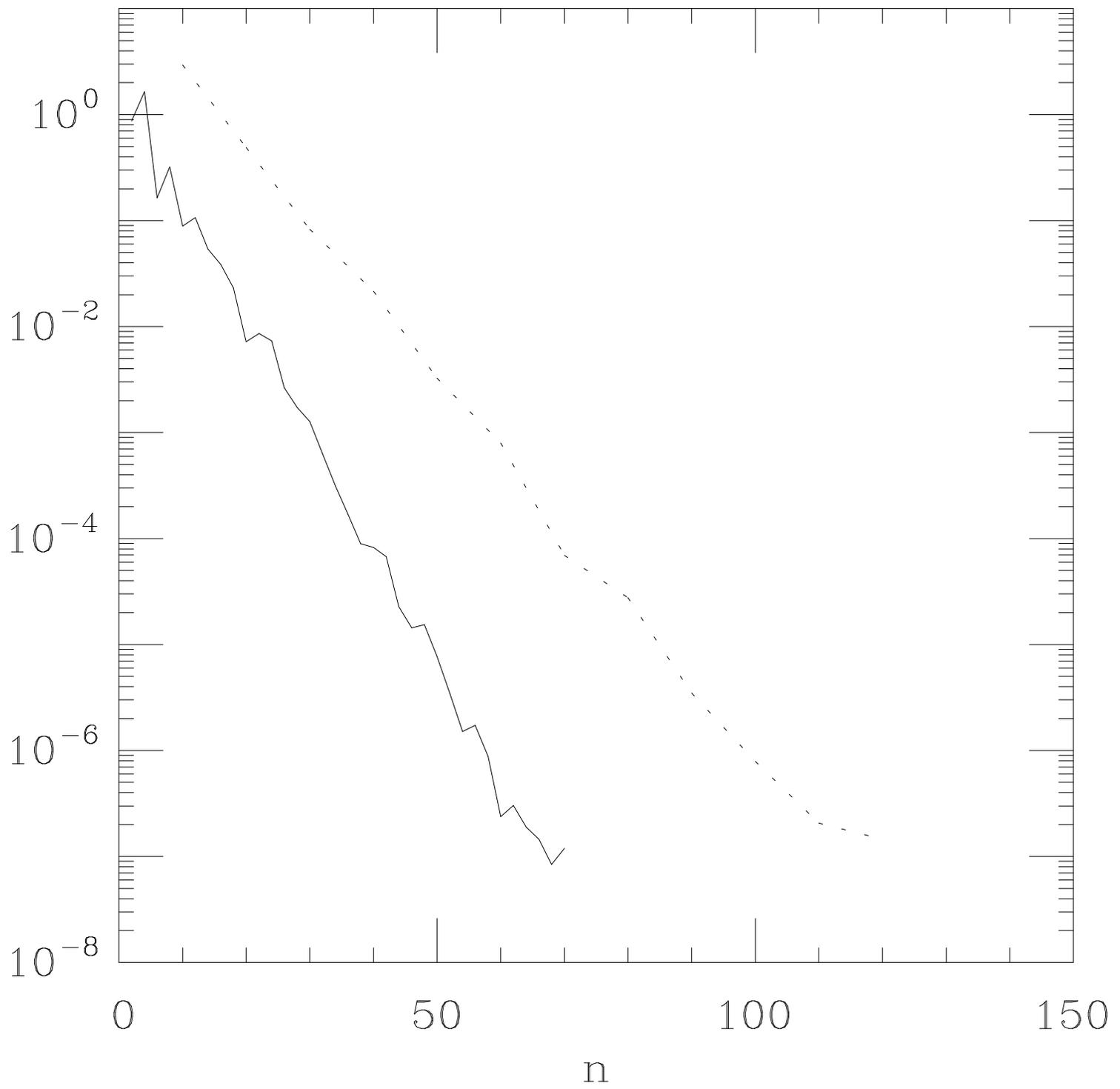
\end{document}